\documentstyle[12pt]{article}

\begin{document}
\begin{center}
{\bf \Large Specific heat  of high temperature superconductors:
Role of $\mid \psi \mid^{4}$ term in the Ginzburg-Landau free energy}\\
\vspace*{0.3 true in}
{\large B. C. Gupta$^\dagger$ and K. K. Nanda$^{\ddagger *}$}\\
{\it Institute of Physics, Bhubaneswar-751005, India}\\
\vspace*{0.3 true in}
\end{center}
\begin{abstract}
We have derived the expression for the specific heat by using 
Ginzburg-Landau (GL) theory by taking $\mid \psi \mid^{4}$ into 
account in the Hartree approximation. Without this term, the 
specific heat diverges at $T=T_{c}(B)$. It is also shown that 
width and shape of the transition depends on the value of $\alpha$
and $\beta$, the coefficients in the GL free energy.
\end{abstract}
\vspace{0.5in}

\vspace{1in}
\noindent
$^{\dagger}$ e-mail: bikash@iopb.ernet.in \\
\noindent
$^{\ddagger}$ e-mail: nanda@iopb.ernet.in \\
\noindent
$^{*}$ Corresponding author
\newpage
\section{\bf Introduction}

The most interesting feature of high temperature superconductors 
(HTSCs) is that
the superconducting transition is broadened in presence of magnetic field
which has been understood in terms of thermodynamic fluctuations near the
superconducting phase transition and has drawn a great interest and many
theories have been developed \cite{ike,ull,ztc,moor,over}. It is
expected that fluctuations of
the normal phase into the superconducting phase above the transition
temperature $T_{c}$ can give rise 
to additional contributions to both thermodynamic and transport quantities.
These fluctuations can be viewed as superconducting droplets
spontaneously appear and
disappear on time scales of $\hbar /k_{B} \mid T-T_{c} \mid$ \cite{tink}. Our
discussion is based on the lowest landau level approximation. According to
this approximation, the applied field forces the superconducting electrons to
move in the lowest landau orbitals perpendicular to the field, thereby,
reducing the effective dimensionality of the system. Within this 
approximation, it is shown \cite{ztc} that physical properties
exhibit scaling with scaling variable
$x=[\frac{T-T_{c}(B)}{(TB)^{n}}]$, where $n$=2/3 for 3-dimensional 
(3D) systems and 1/2 for 2D systems that has been observed experimentally 
\cite{welp}.\\

\par
If the quartic term is dropped in the GL free energy, an analytic expression
for the specific heat is obtained which diverges at $T=T_{c}(B)$ 
\cite{thou}. Close to $T_{c}(B)$, this term plays an important role
and cannot be ignored. The effect of including the quartic term for
layered superconductors is studied { \it implicitly} by Quader and Abrahams 
\cite{qua}.
But, if the variation of order parameter along the direction of the applied
field is smooth enough, then the problem is reduced to ordinary 3D
GL equation \cite{tink2}. Sufficiently near $T_{c}$, the coherence length
along the $c-$axis $\xi_{c}$ will be so large that $\xi_{c} > s$, the
interplanar distance and the 3D description of GL theory is justified.

\par 

An exact evaluation of the partition function Z is not possible in the
presence of the quartic term. However, the partition function can be evaluated
within a Hartree-type approximation where $\mid \psi \mid^{4}$ is replaced
by 2$< \mid \psi \mid^{2} > \mid \psi \mid^{2}$ \cite{mask}.
In this paper, we take the quartic term into account and derive the specific
heat in the Hartree approximation.

\section{\bf Hartree Approximation}

The free energy functional for a superconductor in a uniform flux density
$B$ near the $T_{c}$ has the form \cite{thou}
\begin{equation}
F_{s}=F_{n}+\int dV [\frac{1}{2m} \mid(-{\it i}\hbar \nabla - eB
\times r) \psi \mid^2 + \alpha \mid \psi \mid^{2} + \frac{\beta}{2} 
\mid \psi \mid^{4} +.....]
\end{equation}
\noindent
where $\alpha= \alpha_{0} t$ and $\beta=\beta_{0}$.

To calculate the fluctuations above $T_{c}(B)$, we can use the free energy 
functional of Eq. (1).
The terms quadratic in $\psi$ is considered which gives the
specific heat to be divergent at $T=T_{c}(B)$. 
Close to $T_{c}(B)$, the terms quartic in the fluctuations can not be 
ignored. Within the Hartree approximation,
\begin{equation}
\Delta F[\psi]=F_{s}-F_{n}=\int dV [\alpha \mid \psi \mid^{2}
+\beta < \mid \psi \mid^{2} > \mid \psi \mid^{2}
+ \frac{1}{2m} \mid(-{\it i}\hbar \nabla - eB \times r) \psi \mid^2 ]
\end{equation}
\noindent
In the free energy expression, we have dropped the term
$B^{2}/2\mu_{0}$ because $B$ is taken to be the mean value of the
induction and neglects the fluctuations. The free energy can be 
diagonalized in terms of the solutions of the Eq. 
\begin{equation}
\frac{1}{2m} (-{\it i}\hbar \nabla - eB \times r)^{2} \psi + \alpha \psi 
+ \beta < \mid \psi \mid^2 >= E \psi
\end{equation}

Writing $\psi=\sum_{q} C_{q} \psi_{q}(r)$, we get the eigen values are
\begin{equation}
E_{q}=\alpha+(2n+1)\frac{e \hbar B}{m}+\frac{\hbar^2 k_{z}^2}{2m}
+\beta < \mid \psi \mid^{2} >=
\alpha_{B}+2n \frac{e \hbar B}{m}+\frac{\hbar^2 k_{z}^2}{2m}
+\beta < \mid \psi \mid^{2} > 
\end{equation}
with degeneracy $eB/\pi \hbar$ per unit area and $\alpha_{B}=
\alpha+\frac{e \hbar B}{m}$.

In the vicinity of the upper critical field, the dominant contribution 
comes from the lowest landau level ($n$=0). 
The order parameter averages are given by \cite{mask}
\begin{equation}
< \mid \psi \mid^2 >= k_{B}T \sum_{q} \frac{1}{E_{q}}
\end{equation}
and the specific heat can be calculated as \cite{thou},
\begin{equation}
C=k_{B}T^{2}  \sum_{q} \frac{1}{E_{q}^{2}}(\frac{dE_{q}}{dT})^{2}
\end{equation}

\subsection {\it Specific Heat in 2D Systems}

In case of 2D, for the applied field perpendicular to the film,
the fluctuations are zero dimensional and 
the $k_{z}$ component is suppressed. The average of 
the $\mid \psi \mid^{2}$ for the film of thickness d
can be evaluated as \cite{lee},
\begin{equation}
< \mid \psi \mid^2 >
=\frac{-\alpha_{B}+\sqrt{\alpha_{B}^{2}+(\frac{eB}{d \pi \hbar}) 
\beta k_{B}T}}{2 \beta}
\end{equation}
\noindent 
This along with Eq.(6) gives
\begin{equation}
{C_{2D}}={{\left(\frac{eB}{\pi \hbar}\right) kT^2 \left(\frac{\alpha_{0
}}{2T_c} + \frac{1}{4} \frac{1}{\sqrt{\left( {\alpha_{B}}^2 + 
\beta kT \left( \frac{eB}{d \pi \hbar} \right ) \right )}}
\left( \frac{2 \alpha_{0} \alpha_{B}}{T_c} + 
\beta k \left( \frac{eB}{d \pi \hbar}\right)  \right) \right)}\over{
\frac{1}{4} \left(\alpha_{B} + \sqrt{\left( {\alpha_{B}}^2
+ \beta kT \left( \frac{eB}{d \pi \hbar} \right)
\right)} \right )^2}}
\end{equation}

As $\beta$ is independent of temperature and $\alpha$ decreases
and as $T \rightarrow T_{c}(B)$, $\alpha_{B} \rightarrow 0$
\begin{equation}
< \mid \psi \mid^2 >=(\frac{eB}{\pi \hbar})^{1/2} \beta^{-1/2}
\end{equation}
which is non-zero and give rise to a non divergent specific heat.

\subsection {Specific Heat in 3D Systems}   

Replacing the sum by integration, the average of the $\mid \psi \mid^{2}$ is
obtained as
\begin{equation}
< \mid \psi \mid^2 >= \frac{(\frac{eB}{4 \pi \hbar})k_{B}T 
\sqrt{\frac{m}{2}}}{\sqrt{\alpha_{B}+\beta < \mid \psi \mid^2 >}}
\end{equation}
which gives,
\begin{equation}
< \mid \psi \mid^2 >=(\frac{m}{2})^{2/3}
(\frac{eB}{\pi^2})^{2/3} \beta^{-2/3}
\end{equation}
as $T \rightarrow T_{c}(B)$. Hence the specific heat is non-divergent
and is given by
\begin{equation}
{C_{3D}}={{({eB\over \pi \hbar})kT^2 \left(\frac{\alpha_0}{T_c} +
\frac{2}{3}\left( \sqrt{\frac{m}{2}} \frac{eB}{\pi \hbar} \beta k\right)^
\frac{2}{3} \frac{1}{T^\frac{1}{3}}\right)}\over{\left( \alpha_{B} +
\left(\sqrt{\frac{m}{2}} \left(\frac{eB}{\pi \hbar} \right) kT\right)^
\frac{2}{3}\right)^\frac{3}{2}}}
\end{equation}

\section{\bf Discussion}

The fluctuation in specific heat above $T_{c}(B)$ in case of 3D differs
by a factor $\sqrt{2}$ at a corresponding distance below $T_{c}(B)$
because the coherence length for $\psi(r)$ is smaller by a factor $\sqrt{2}$
below $T_{c}(B)$ in comparison with the region above $T_{c}(B)$
\cite{tink}. Whereas,
the specific heat in case of 2D will be the same and that has been
observed experimentally.

Fig.1 shows the temperature dependence of the specific heat in case
of 2D systems originated from fluctuations for different values of
$\alpha_{0}$ and $\beta_{0}$. It is noted from fig.1 that the width and
the shape of the transition depends on $\alpha_{0}$ and $\beta_{0}$.
Fig.1(a) corresponds to $\alpha_0$ = 10, $\beta_0 = 5000$ and
fig.1(b) corresponds to $\alpha_0$ = 100, $\beta_0 = 10000$. For both
the figures the dashed and the solid curve corresponds to high
(4T) and low (2T) external magnetic fields respectively. For our
calculation we have taken $\frac{dB_{c_{2}}}{dT} = -3.2$. 
It has been shown \cite{kkn1} that the field dependence of the specific heat
below $T_{c}(B)$ can be accounted for in the mean field approximation
by evaluating the
interactions between vortices. Here, the vortex structure has not been
included in these calculations which influence
the specific heat below $T_{c}(B)$. Therefore, below the transition
temperature we do not expect the 
agreement of our result with the experimental data. However, we
note from figure 1(a) and 1(b) that the width of the transition
increases with magnetic field. This is in accordance with the
experiments on both low $T_{c}$ \cite{bar} as well as high-$T_c$
materials \cite{isrg,janod}.
Our result also gives similar behaviour
above $T_{c}(B)$ as obtained by Bray \cite{bray} using screening theory.

\section{\bf Conclusion}

We have presented  and discussed the results for the fluctuation 
specific heat, both with and without the quartic term in the free
energy. An inspection of our results indicate that the interaction
term plays an important role.
The specific heat does not diverge in contrast to the GL theory without
$\mid \psi \mid^4$ term. We have also shown that the decrease of
the sharpness of transition of the fluctuation specific heat with the
increase of magnetic field as is observed in experiment. Above
$T_{c}(B)$, the nature of the fluctuation specific heat curve is also
similar to that, obtained from screening theory, agreeing the
experimental result.

\newpage

\begin{figure} [pt]
\centerline{{\bf Figure Captions}}

\caption{\label {Fig. 1:}} ($\bullet$) 
The temperature dependence of specific heat in case of 2D systems that
originates from fluctuations in different magnetic fields and
different values of $\alpha_0$ and $\beta_0$. a) $\alpha_0 = 10,
\beta_0 = 5000$. b) $\alpha_0 = 100, \beta_0 = 10000$.
\end{figure}
\end{document}